\documentclass[reprint,aps,superscriptaddress,twocolumn,showpacs,longbibliography]{revtex4-2}
\usepackage{amsmath,amsfonts,amssymb,amsthm,graphics,graphicx,epsfig,bbm}
\usepackage[colorlinks=true,citecolor=blue,linkcolor=magenta,urlcolor=blue]{hyperref}
\usepackage[usenames]{color}
\usepackage[dvipsnames]{xcolor}
\usepackage{graphicx}
\usepackage{tikz}
\usepackage{subfigure}
\usepackage{float}
\usepackage{amsmath}

\definecolor{mycolor}{RGB}{106,81,162}

\theoremstyle{plain}
\newtheorem{thm}{Theorem}
\newtheorem{corol}{Corollary}

\theoremstyle{definition}

\theoremstyle{remark}

\renewcommand{\Re}{\mathrm{Re}}
\renewcommand{\Im}{\mathrm{Im}}

\usepackage{mathtools}

\newcommand{\iden}[1]{
    \ifthenelse{\equal{1}{\string #1}}
  {
   \mathbbm{1}
  }
  {
   \mathbbm{1}^{\otimes#1}}
  }
\newcommand{\ketzero}[1]{
    \ifthenelse{\equal{1}{\string #1}}
  {
   \ket{0}
  }
  {
   \ket{0}^{\otimes#1}}
  }
\newcommand{\brazero}[1]{
    \ifthenelse{\equal{1}{\string #1}}
  {
   \bra{0}
  }
  {
   \bra{0}^{\otimes#1}}
  }
\newcommand{\ketone}[1]{
      \ifthenelse{\equal{1}{\string #1}}
    {
     \ket{1}
    }
    {
     \ket{1}^{\otimes#1}}
    }
  \newcommand{\braone}[1]{
      \ifthenelse{\equal{1}{\string #1}}
    {
     \bra{1}
    }
    {
     \bra{1}^{\otimes#1}}
    }

\usepackage{todonotes}
\newcommand{\rev}[1]{{\color{red}#1}}

\usepackage{physics}  
\usepackage{dsfont}

\begin{document}

\title{Mixed Quantum-Semiclassical Simulation}

\author{Javier Gonzalez-Conde} 
\affiliation{Department of Physical Chemistry, University of the Basque Country UPV/EHU, Apartado 644, 48080 Bilbao, Spain}
\affiliation{EHU Quantum Center, University of the Basque Country UPV/EHU, Apartado 644, 48080 Bilbao, Spain}



\author{Andrew T. Sornborger} 
\affiliation{Information Sciences, Los Alamos National Laboratory, Los Alamos, NM, USA.}

\date{\today}

\begin{abstract}
\noindent
We study the quantum simulation of mixed quantum-semiclassical (MQS) systems, of fundamental interest in many areas of physics, such as molecular scattering and gravitational backreaction. A basic question for these systems is whether quantum algorithms of MQS systems would be valuable at all, when one could instead study the full quantum-quantum system. We study MQS simulations in the context where a semiclassical system is encoded in a Koopman-von Neumann (KvN) Hamiltonian and a standard quantum Hamiltonian describes the quantum system. In this case, because KvN and quantum Hamiltonians are constructed with the same operators on a Hilbert space, standard theorems guaranteeing simulation efficiency apply.
We show that, in this context, {\it many-body} MQS particle simulations give only nominal improvements in qubit resources over quantum-quantum simulations due to logarithmic scaling in the ratio, $S_q/S_c$, of actions between quantum and semiclassical systems. However, {\it field} simulations can give improvements proportional to the ratio of quantum to semiclassical actions, $S_q/S_c$. Of particular note, due to the ratio $S_q/S_c \sim 10^{-18}$ of particle and gravitational fields, this approach could be important for semiclassical gravity. We demonstrate our approach in a model of gravitational interaction, where a harmonic oscillator mediates the interaction between two spins. In particular, we demonstrate a lack of distillable entanglement generation between spins due to classical mediators, a distinct difference in dynamics relative to the fully quantum case.
\end{abstract}

\maketitle

\section{Introduction}

\begin{figure*}
\centering 
\includegraphics[width=2.05\columnwidth]{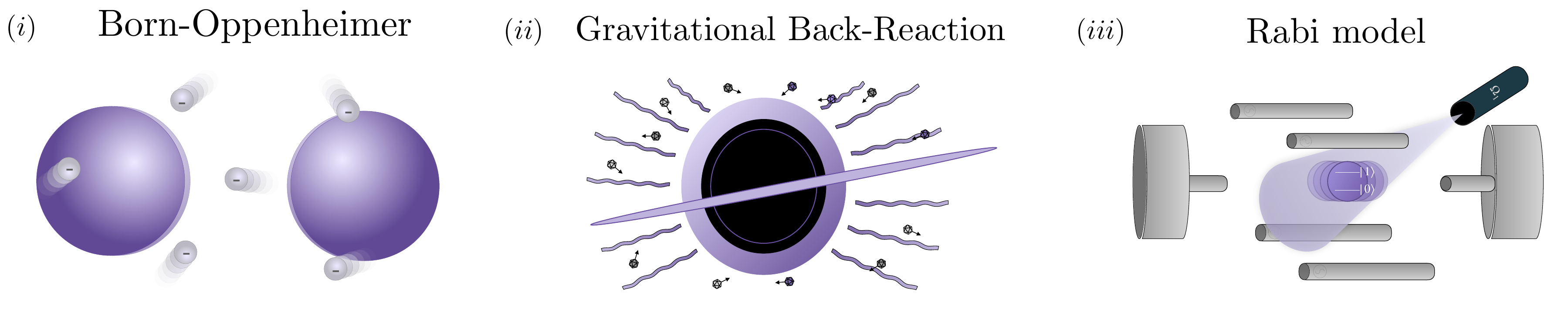}
\caption{Examples of MQS systems of interest. (i) In molecular scattering, the Born-Oppenheimer approximation is used to describe the light electronic degrees of freedom quantum mechanically, but the heavy nuclear system semiclassically \cite{combes1981born, woolley1977molecular,essen1977physics, simon2023improved}.  (ii) Backreaction in gravity is studied with various semiclassical gravitational approaches \cite{Blanco_Pillado_2019, Schander_2021, Blanco_Pillado_2019, Schander_2021, hu1995back, Tsamis_2014,hu1995back, padmanabhan1989semiclassical, preskill1992black, banerjee2008quantum, Bose_2017, Bose_2020,Marletto_2017,Carney_2021, Carney_2022, Danielson_2022,Ma_2022, PhysRevLett.128.110401}. (iii) The Jaynes-Cummings model serves as a toy model for many MQS systems \cite{shore1993jaynes, babelon2009semi,alscher2001semiclassical, braak2016semi}.}
\label{fig:examples}
\end{figure*} 

Mixed quantum-semiclassical (MQS) systems are studied in many areas of physics \cite{combes1981born, woolley1977molecular,essen1977physics,  simon2023improved, Blanco_Pillado_2019, Schander_2021, Blanco_Pillado_2019, Schander_2021, hu1995back, Tsamis_2014, hu1995back, padmanabhan1989semiclassical, preskill1992black, banerjee2008quantum, Bose_2017, Bose_2020,Marletto_2017,Carney_2021, Carney_2022, Danielson_2022,Ma_2022, PhysRevLett.128.110401, shore1993jaynes, babelon2009semi, alscher2001semiclassical, braak2016semi}. In molecular scattering, the Born-Oppenheimer approximation, Fig.~\ref{fig:examples}~(i), describes the light electronic degrees of freedom quantum mechanically, but the heavy nuclear system semiclassically \cite{combes1981born, woolley1977molecular,essen1977physics,  simon2023improved}.  Backreaction in gravity, Fig.~ \ref{fig:examples}~(ii), is studied with various semiclassical gravitational approaches \cite{Blanco_Pillado_2019, Schander_2021, Blanco_Pillado_2019, Schander_2021, hu1995back, Tsamis_2014,hu1995back, padmanabhan1989semiclassical, preskill1992black, banerjee2008quantum, Bose_2017, Bose_2020,Marletto_2017,Carney_2021, Carney_2022, Danielson_2022,Ma_2022, PhysRevLett.128.110401}. And the Jaynes-Cummings model, Fig.~\ref{fig:examples}~(iii), serves as a toy model for many MQS systems \cite{shore1993jaynes, babelon2009semi,alscher2001semiclassical, braak2016semi}.

One of the main problems in this area of research concerns the study of backreaction between classical and quantum theories, which is of particular interest, since quantum fields can generate entanglement, whereas classical fields do not. Typical semiclassical derivations from field theory result in a standard coupling \cite{born1985quantentheorie}. However, some researchers, particularly in gravitational theory, which is non-renormalizable, have suggested that MQS theories could be fundamental, giving rise to open questions involving how to couple quantum with semiclassical fields and giving rise to suggestions of non-standard couplings \cite{Wilczek2015NotesOK}.


A question that arises naturally from the study of these systems is whether MQS theory could benefit from quantum simulation algorithms, as a significant component of these models is quantum, and hence costly to compute classically \cite{Lloyd_1996, doi:10.1137/S0097539704445226, ref3, 10.1145/1993636.1993682, Tacchino_2019}. In this regard, algorithms for Hamiltonian simulation on quantum computers have reached a mature stage \cite{Lloyd_1996, Preskill_2018, Tacchino_2019, Georgescu_2014, Barends_2016, Camps_2022, Trotter1959OnTP, Clinton_2021, Efekan_2022, Berry_2006,Low_2019, Childs2012, suzuki1976generalized, sornborger1999higher, Pastori_2022, dong2022quantum, Campbell_2019, Layden2022, Watkins_2022, Berry_2020, Chen_2021, Berry_2015, yordanov2020efficient}. Quantum simulations are known to use logarithmically fewer resources versus their equivalents on classical computers \cite{Berry_2006, yordanov2020efficient}. Complexity bounds are now well-understood and current quantum simulation algorithms are at the theoretical limit for accuracy \cite{Clinton_2021, Preskill_2018, Layden2022}.

Additionally, quantum algorithms for solving differential equations and performing classical simulations are of recent interest. A range of quantum algorithmic approaches for classical simulations has recently been studied, including Carleman \cite{forets2017explicit,Liu_2021, an2022efficient,lewis2023limitations} and Koopman-von Neumann methods \cite{Koopman,PhysRevA.105.052404,Gay_Balmaz_2022, bondar2019koopman, Wilczek2015NotesOK, lin2022koopman, PhysRevResearch.2.043102,JIN2023112149,simon2023improved} for lifting finite dimensional, classical nonlinear differential equations to a linear description in an infinite dimensional Hilbert space. Recent results show an exponential advantage when simulating classical systems of harmonic oscillators \cite{babbush2023exponential}.

It is unclear, however, if an exponential advantage may be obtained, generally, for simulating classical systems  \cite{lin2022koopman,babbush2023exponential}. And, in some cases, it is known that it cannot \cite{lewis2023limitations}. One reason for this is that although quantum simulations of quantum systems overcome the `sign problem' in quantum Monte Carlo caused by quantum interference  \cite{troyer2005computational,pan2022sign}, this problem does not exist for classical Monte Carlo and therefore classical Monte Carlo methods are already very competitive. Therefore, it can be difficult to find quantum algorithms that give better than a polynomial advantage for classical simulation. 

In the case of quantum algorithms for MQS simulations, an opportunity exists to make use of a quantum simulation of a classical system by coupling it with a quantum simulation of a quantum system. In this case, a computational advantage is obtained, generally, since the quantum part of the system becomes exponentially efficient, even if the classical part does not. Therefore, with MQS simulations, we have the opportunity to study backreaction \cite{Blanco_Pillado_2019, Schander_2021, hu1995back, Tsamis_2014}, among other phenomena, with exponential efficiency on a quantum computer.

In this article, we construct an MQS simulation algorithm with a semiclassical subsystem described using a Koopman-von Neumann (KvN) framework within an otherwise quantum simulation. KvN is most amenable to MQS simulation in the Schr\"odinger picture since semiclassical state evolution is governed by a quantum Hamiltonian on a Hilbert space. The defining difference in combined KvN and quantum Hamiltonian simulation is that the KvN simulation involves system operators that commute with each other. Thus, MQS simulation algorithms based on KvN are efficient and use logarithmically fewer resources than the equivalent classical algorithm. Additionally, for MQS systems, our algorithm can use fewer resources than the equivalent quantum-quantum algorithm. We show that although this reduction in resources is nominal for MQS many-body simulations, it can be significant for MQS field simulations \cite{jordan2012quantum}. Finally, we implement our approach with a proof-of-principle model of two, two-level quantum systems with an interaction mediated by a harmonic oscillator. We compare simulation of the model with both quantum and classical oscillators and study and contrast their entanglement properties. In particular, we show that the MQS system does not generate distillable entanglement, whereas the fully quantum system does.

\section{Semiclassical Dynamics}

 The purpose of semiclassical mechanics is to obtain a formalism that enables us to express the dynamics of classical (Liouville) systems with quantum operators. In quantum mechanics, we assume that the dynamics of the quantum density matrix of a closed system, $\hat{\rho}$, is governed by the von Neumann equation 
\begin{equation} 
   \partial_t \hat{\rho} = -\frac{i}{\hbar} [\hat{H}, \hat{\rho}] \;,
   \label{denseq}
\end{equation}
where $\hat{H}$ is the system Hamiltonian and $[ \hat{H}, \hat{\rho}]=~\hat{H} \hat{\rho}- \hat{\rho} \hat{H}$ the commutator. On the other hand, in the context of classical mechanics, the time evolution of a phase space distribution function, $f(p,q)$, is described by the Liouville equation,

\begin{equation} 
  \frac{\partial f}{\partial t}=-\{\,f,H\,\} 
   \label{denseq}
\end{equation}
with $\{\,f,H\,\}= \sum_{i=1}^N\left(\frac{\partial f}{\partial q_i}\frac{\partial H }{\partial p_i}
-\frac{\partial f}{\partial p_i}\frac{\partial H}{\partial q_i}\right)$, the Poisson bracket.




We aim to put quantum and semiclassical systems on the same footing in order to make use of quantum Hamiltonian simulation methods. In this regard, we use phase-space quantum mechanics \cite{GROENEWOLD1946405, moyal_1949} to derive a classical Liouville equation, then construct a Hamiltonian using Koopman-von Neumann mechanics \cite{Koopman, PhysRevA.105.052404,Gay_Balmaz_2022, Bondar_2019, Wilczek2015NotesOK}.

Phase-space quantum mechanics relies on transforming the quantum density matrix, $\hat{\rho}$, into a \textit{pseudo}-distribution, $\gamma({q},{p})$, in phase space. The standard method to achieve this mapping is via the Wigner transform \cite{heller1976wigner}, 

\begin{equation}
   \gamma_\omega({q},{p}) = \frac{1}{(2\pi\hbar)^2} \int_{\mathbb{R}^d} \rho ( {q} - \frac{1}{2} {r}, {q} + \frac{1}{2} {r} ) e^{i{p}\cdot{r}/\hbar} d{r} \; ,
\end{equation}
where
\begin{equation}
  \rho ( {q} - \frac{1}{2} {r}, {q} + \frac{1}{2} {r} ) \equiv \phi ( {q} - \frac{1}{2} {r} ) \phi^* ( {q} + \frac{1}{2} {r} ) \; .
\end{equation}
This transformation is equivalent to the standard Schr\"odinger formalism, but the transform takes the density matrix into a quasiprobability distribution in phase space \cite{PhysRevLett.10.277, Rundle_2021}.

Following this definition, the quasiprobability distribution, $\gamma_w$, using (\ref{denseq}), obeys the dynamical equation
\begin{equation}
  \partial_t \gamma_w = -\frac{2}{\hbar} H \sin\left(\frac{\hbar \Lambda}{2}\right) \gamma_w \; ,
  \label{gammaeq}
\end{equation}
where $H$ is the classical Hamiltonian and the differential operator $\Lambda$ is defined
\begin{equation}
   \Lambda \equiv \overleftarrow{\partial}_p \overrightarrow{\partial}_q - \overleftarrow{\partial}_q \overrightarrow{\partial}_p \; .
\end{equation}
We expand (\ref{gammaeq}) to zeroth order in $\hbar$, resulting in a semiclassical approximation of the dynamics given by the classical Liouville equation \cite{heller1976wigner}
\begin{equation}
   \partial_t \gamma_w({p},{q}) = 
   -\frac{2}{\hbar} H \left(\frac{\hbar \Lambda}{2}\right) \gamma_w + O(\hbar^2) \nonumber
\end{equation}
giving
\begin{equation}
   \partial_t \gamma_c({p},{q}) \approx -\partial_p H \; \partial_q \gamma_c + \partial_q H \; \partial_p \gamma_c\; ,
   \label{liouville}
\end{equation}
where $\gamma_c$ denotes semiclassical evolution in the phase space of the pseudo-distribution $\gamma_w$. 
It is important to realize that, although $\gamma_c$ is a pseudo-distribution, for any measurement probability, $P_A$, where $A = \ket{a}\bra{a}$, and $\ket{a}$ is a quantum state, $P_A = \mathrm{Tr}\{\rho A\} = \mathrm{Tr}\{\gamma_w A_w\}$, and $A_w$ is the Wigner transform of $A$. Therefore, there is no possibility of measuring negative probabilities. 

Note that when we take the limit $\hbar \rightarrow 0$, we really are using a shorthand for taking a limit $\hbar/S_0 \rightarrow 0$, where $S_0$ is a characteristic action, or a variable relevant to the action, in the theory.



\section{Koopman-von Neumann Hamiltonian Dynamics}\label{KvN}




In the Liouville equation, (\ref{liouville}), derived above for semiclassical mechanics, Hamiltonian dynamics governs motion in phase space. Defining the square-root of the pseudo-distribution as $\gamma_c \equiv ~ \psi(x,p)^* \psi(x,p)$, where $\psi(x,p)$ is a complex function, it is possible to derive an equation for a wavefunction, $\psi$,
\begin{equation}
   i \partial_t \psi = \hat{L} \psi \; ,
\end{equation}
where $\hat{L}$ is the Hermitian operator
\begin{equation}
   \hat{L} = i \left( -\partial_p H \; \partial_q\psi + \partial_q H \; \partial_p \psi \right) \; .
\end{equation}
Note the close resemblance to a quantum theory here, since $\hat{L}$ is Hermitian. However, some work remains to obtain a representation that fully obeys the axioms of Quantum Mechanics.

Our goal is to develop an operator representation of the phase space dynamics of $\psi(q,p)$ in a Hilbert space \cite{Wilczek2015NotesOK,Koopman}. This will give us a quantum, Schr\"odinger representation of classical Hamiltonian dynamics that may be implemented with quantum registers and methods from quantum simulation on a quantum computer, while still maintaining semiclassical dynamics. We note that nonlinear Hamiltonian systems are readily represented with this formalism, called Koopman-von Neumann dynamics. 

We introduce axioms that are standard in quantum theory, except that for the classical theory, we require that the commutator
\begin{equation}
   [\hat{q}, \hat{p}] = 0 \; .
\end{equation}
We note that the classical commutator may formally be considered to be the limit as $\hbar \rightarrow 0$ of the quantum commutator
\begin{equation}
   [\hat{q}, \hat{p}] = i\hbar \; .
\end{equation}

The axioms that we will use in our framework \cite{Wilczek2015NotesOK} are: (1) The wavefunction, $\Psi$, is normalized as $\langle \Psi(t) | \Psi(t) \rangle = 1$; (2) The expectation value of an observable $\hat{A}$ is given by $\mathrm{Tr} \{\rho A\} = \langle \Psi(t) | \hat{A} | \Psi(t) \rangle$; (3) The probability of measuring a value, $a$, of an observable is $\mathrm{Pr}(a) = | \Pi_a |\Psi(t)\rangle |^2$, with $\Pi_a$ a projection operator on the space defined by $\hat{A}$ with eigenvalue $a$; (4) The composition of two systems, $A$ and $B$, is given by the tensor product of their Hilbert spaces, $\Psi_{AB} \equiv \Psi_A \otimes \Psi_B$.

In this formalism, in order to preserve probability (i.e. axiom (1) above), evolution in time is governed by a unitary operator, $U(t)$. However, note the important fact that $\hat{L}$ is {\it not} the generating operator for time evolution. It turns out that, in order for $\hat{q}$ and $\hat{p}$ to commute, two new operators, $\hat{\lambda}_q$ and $\hat{\lambda}_p$, must be introduced. We call these ghost momenta. In this formalism, both $\hat{q}$ and $\hat{p}$ can be considered as configuration space operators in a quantum theory, and, as such, they commute. The ghost momenta, $\hat{\lambda}$, {\it do not} commute with their corresponding configuration operators:
\begin{eqnarray}
    [ \hat{q}, \hat{\lambda}_q ] & = & i \cr
    [ \hat{p}, \hat{\lambda}_p ] & = & i \; .
\end{eqnarray}
So, in this theory, the size of the Hilbert space must be doubled to accommodate the commuting configuration space operators, $\hat{q}$ and $\hat{p}$.

\begin{figure}[t!]
\centering 
\includegraphics[width=1\columnwidth]{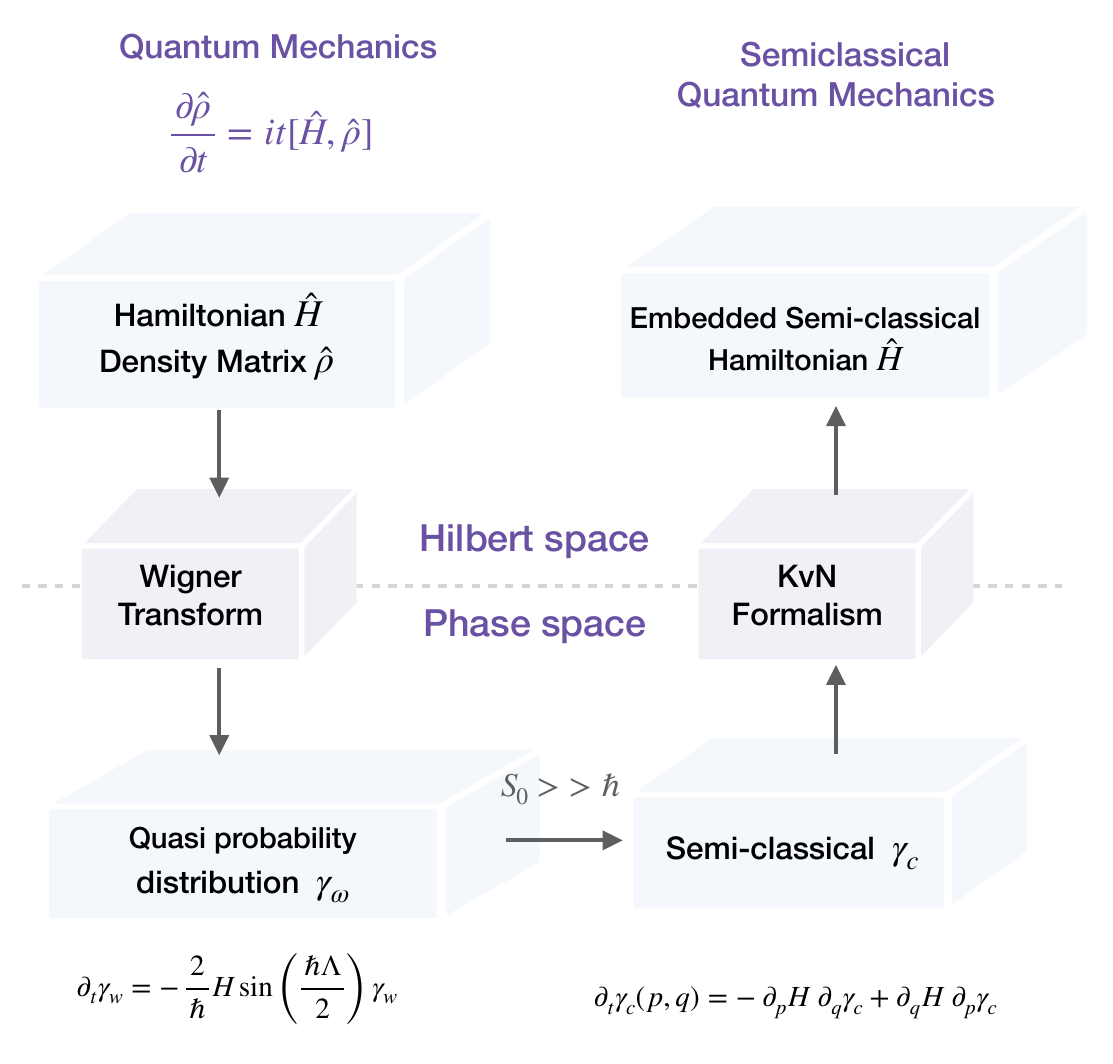}
\caption{Scheme followed to obtain a quantum representation of semiclassical-dynamics. First, we transform the state $\hat{\rho}$, and Hamiltonian of our system $\hat{H}$ to phase space via the Wigner Transform. In phase space, we obtain a dynamical equation associated with a quasiprobability distribution, $\gamma_w$. In order to move to the semiclassical framework we take the limit of  the characteristic action $S_0$, to be much larger than $\hbar$, which brings us to the approximate dynamics of the semiclassical quasiprobability distribution, $\gamma_c$. 
Then, with a KvN formulation, we rewrite the semiclassical Liouvillian dynamics in the Schr\"odinger formulation in order to be able to implement the semiclassical dynamics on a quantum computer.}
\label{fig:schame}
\end{figure}

With the introduction of ghost momenta, we can now write down the generator of time evolution for this theory,
\begin{equation}
   \hat{H}_\mathrm{KvN} \equiv \widehat{\partial_p H} \hat{\lambda}_q - \widehat{\partial_q H} \hat{\lambda}_p \; ,
   \label{eq:KvNHamiltonian}
\end{equation}
where $H$ is the classical Hamiltonian and the hat over the classical Hamiltonian derivative terms represents the promotion of phase space variables, $(q,p)$, to operators $(\hat{q},\hat{p})$. $\hat{H}_\mathrm{KvN}$ gives the time evolution of the semiclassical system on Hilbert space, 
\begin{equation}
| \Psi(t) \rangle = e^{{-i \hat{H}_\mathrm{KvN} t}} | \Psi(0) \rangle \; .
\end{equation}
This approach is depicted in Fig. \ref{fig:schame}.

\section{Quantum-Semiclassical Dynamics}


In the KvN dynamics described in Sec.~\ref{KvN}, we ignored the global phase of the KvN wavefunction since it did not contribute to any observables and had no impact on the classical dynamics. However, when we couple quantum and semiclassical systems, we can no longer ignore the semiclassical phase, since the dynamics now becomes sensitive to it \cite{bondar2019koopman}.

Retaining the phase, $\hat{W}$, in KvN dynamics gives us the KvN Hamiltonian
\begin{equation}
\label{eq:kvn_formalism}
     \hat{H}_\mathrm{KvN} \equiv \widehat{\partial_p H} \hat{\lambda}_q - \widehat{\partial_q H} \hat{\lambda}_p + \hat{W} \; ,
\end{equation}
where $\hat{W}$ generates a phase on the semiclassical Hilbert space.

Kostant \cite{kostant1972linebundles}, and later recapped in Bondar, et al. \cite{bondar2019koopman} and Joseph \cite{PhysRevResearch.2.043102}, note that without the phase, $\hat{W}$, the KvN Lagrangian (used in a variational formulation of KvN dynamics) is not covariant with respect to local phase transformations, $\Psi(q,p) \rightarrow e^{i\varphi(q,p)} \Psi(q,p)$. They further note that this may be overcome by considering a minimally coupled gauge theoretic approach, where a $U(1)$ gauge-covariant Liouvillian is constructed with the replacements
\begin{equation}
   i\hbar \partial_t \rightarrow i\hbar\partial_t - \Phi \;\; \mathrm{and} \;\; i\hbar\nabla \rightarrow i\hbar\nabla + \mathcal{A} \; ,
\end{equation}
where $\Phi = H$ is the gauge potential, and $\mathcal{A} \cdot dz$ is the symplectic potential (with $z = (q, p))$. In this paper, we use the harmonic oscillator gauge
\begin{equation}
   \mathcal{A} \cdot z = \frac{1}{2} \left(p \cdot dq - q \cdot dp \right) \; ,
\end{equation}
with which the phase of the covariant Liouvillian is eliminated for the homogeneous Hamiltonians with quadratic coupling that we consider in our examples \cite{bondar2019koopman}.

A variety of approaches to coupling quantum and semiclassical systems have been considered \cite{bondar2019koopman,gay2021koopman}. Typically, from a fully quantum system, components are identified with parameters that allow one or more components to be made semiclassical in some limit. 

For some systems, such as gravity coupled with quantum fields, it has even been considered that gravity is {\it fundamentally} classical \cite{oppenheim2022constraints,Wilczek2015NotesOK}. Additionally, more formal approaches have been considered, where in an initially classical theory, one system is promoted to a quantum system via canonical quantization, and the other system is (conceptually) separately promoted to a Liouvillian system encoded in a KvN Hamiltonian \cite{bondar2012operational}. 

The general form of the quantum Hamiltonian that we use for these quantum-semiclassical systems is
\begin{equation}
   \hat{H}_\mathrm{qc} = \hat{H}_\mathrm{KvN} + \hat{H}_\mathrm{int} + \hat{H}_\mathrm{q} \; .
   \label{eq:FullHamiltonian}
\end{equation}
where the semiclassical term $\hat{H}_\mathrm{int}$ is derived according to the KvN formalism, Eq.~(\ref{eq:kvn_formalism}), from the fully quantum Hamiltonian. We illustrate this with an example in section \ref{theoretical_model}.

For this quantum-semiclassical system, the quantum density matrix is positive definite. However, the Liouville density, $\gamma_c(p,q)$, which is derived from a Wigner transform can become negative for general quantum-semiclassical interactions. This indicates that the semiclassical subsystem can retain some quantum characteristics. One means of understanding this is given in \cite{heller1976wigner}. Alternatively, the derivation of MQS theories may be studied from an effective field theory approach \cite{brambilla2018born}.




   \label{eq:PointerConstraint}





\section{Dynamical Quantum-Semiclassical Simulations are Efficient}

Quantum signal processing (QSP) methods currently provide optimal quantum algorithms for quantum simulation \cite{Low_2017,gilyen2019quantum}. The following theorem outlines bounds on the time-interval, error, and success probability for QSP methods.
\begin{thm}[Optimal sparse Hamiltonian simulation using quantum signal processing \cite{Low_2017}]\label{optsim}
A $d$-sparse Hamiltonian, $\hat{H}$, on $n$ qubits with matrix elements specified to $m$ bits of precision can be simulated for time-interval $t$, error $\epsilon$, and success probability at least $1 - 2\epsilon$
with $O\left( td\| \hat{H}\|_\mathrm{max} + \frac{\log(1/\epsilon)}{\log \log(1/\epsilon)}\right)$ queries and a
factor of $O[n +~ m\, \mathrm{polylog}(m)]$ additional quantum gates. The quantum simulation is valid for $td\| \hat{H}\|_\mathrm{max} =~ O[\log(1/\epsilon)/\log\log(1/\epsilon)]$. 
\end{thm} 

Since the combined KvN, interaction, and quantum Hamiltonians that we have constructed for the MQS systems considered here consist of a fully quantum operator acting on a Hilbert space that may be encoded in a quantum computer, we have,
\begin{corol}[Optimal Mixed Quantum-Classical Simulation] Simulations of d-sparse interacting quantum and KvN-embedded semiclassical systems are optimal in the sense of Thm.~\ref{optsim}.
\label{cor:OptimalMQSSim}
\end{corol}

We consider MQS simulations that are governed by a $d$-sparse Hamiltonian $\hat{H} = \hat{H}_\mathrm{KvN} + \hat{H}_\mathrm{int} + \hat{H}_\mathrm{Q}$. Here, KvN and interaction Hamiltonians are built from the same operators as the quantum theory. The only difference is the introduction of the $\hat\lambda$ operators, which are standard quantum momentum operators and may be implemented efficiently. Thus, the full Hamiltonian inherits the optimality of quantum simulation using quantum signal processing techniques.

\section{Resource Savings with Quantum-Semiclassical vs. Quantum-Quantum Dynamics} 
\begin{figure}[t!]
\centering 
\includegraphics[width=1\columnwidth]{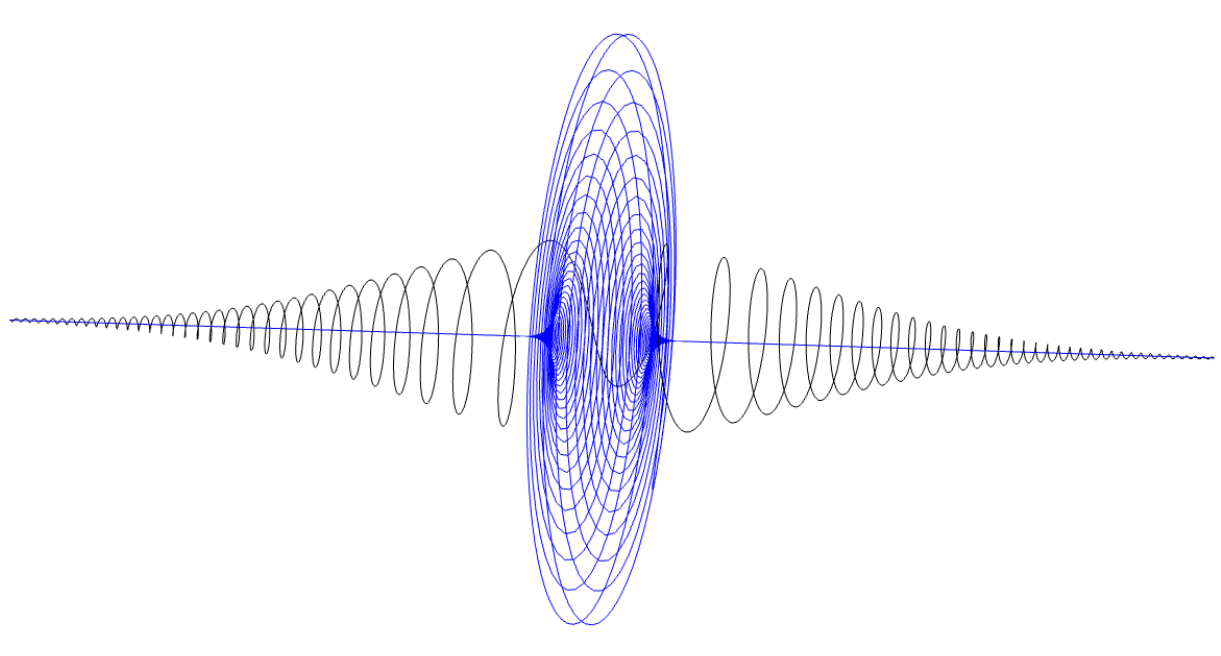}
\caption{Visualization of wavepackets induced by differing actions. This figure gives a geometrical depiction of the difference in wavepacket shapes. A free particle wavepacket with a mass of $10$ units (blue) is plotted with a wavepacket with a mass of $0.1$ units (black). When the ratio between quantum and classical actions, $S_q/S_c$, is large, the particle corresponding to the large action ($S_c$, blue) can be treated semiclassically.}
\label{fig:wavepacket}
\end{figure}

In a theory describing two systems with drastically different effective actions (depending on disparate masses, momenta, or potentials), MQS descriptions can simplify computations, both analytically and numerically, relative to quantum-quantum descriptions. For instance, in Born-Oppenheimer theory describing electron-nucleus interactions, the parameter used to distinguish between quantum electronic degrees of freedom and semiclassical nuclear degrees of freedom is $\sqrt{m_e/m_n}$, where $m_e$ is the electron mass, and $m_n$ is an appropriate nuclear mass scale, typically more than three orders of magnitude larger than the electron mass \cite{combes1981born, woolley1977molecular,essen1977physics, simon2023improved}.

In this limit, the semiclassical field is described by a high frequency wavepacket, where the effective nuclear action is very large compared to the electromagnetic action, which is described by a low frequency wavepacket. This situation (see Fig.~\ref{fig:wavepacket}) will occur whenever the characteristic action of one (semiclassical) system, $S_c$, is significantly larger than that of another. This will be relevant for understanding quantum computational resource savings that give an advantage for MQS systems.

As a specific example, a (dispersive) plane wave describing a free particle of mass, $m$, may be described by the expression
\begin{equation}
   \psi(x,t) = \frac{1}{\sqrt{2\pi\hbar}} e^{ip (x - \frac{p}{2m}t)/\hbar} \; .
\end{equation}
Here, the spatial wavenumber is $p = \sqrt{2mE}$. To numerically describe this particle wave function for a fully quantum simulation, we require a grid with wavenumber $2\sqrt{2mE}$ to resolve the high frequency oscillations due to the particle mass, $m$. This corresponds to the Shannon-Nyquist frequency \cite{shannon1949communication}, whereas for a numerical simulation of the semiclassical equation, the high-frequency quantum wave packet structure does not contribute since it is coarse-grained in the semiclassical approximation. Thus, for an MQS simulation of a quantum particle with mass $m_q$ and a semiclassical particle with mass $m_c$, we only need to resolve the quantum particle, which is lighter and induces lower frequency power in the wave function. In this example, we save, of order $\sqrt{m_q/m_c}$ grid points, assuming some maximum energy scale, $E_\mathrm{max}$.

More generally, we have wavefunctions of the form
\begin{equation}
  \psi(x,t)_{q,c} = A(x,t)_{q,c} e^{i S(x,t)_{q,c}/\hbar} \; .
\end{equation}
where $A$ is a complex amplitude. In this case, we save, of order $S_q/S_c$ grid points with an MQS simulation.

There are two different cases that must be considered to understand the quantum computational resources needed for such simulations. In the first case, consider a many-body simulation. Here, each quantum and semiclassical particle's amplitude is defined on a semiclassical spatial grid of size, $N^\mathrm{semi}_\mathrm{grid}$, which is binarily encoded in a quantum register. Letting $n_\mathrm{q,c}$ represent the numbers of particles simulated for quantum and semiclassical systems (i.e. particles with small or large action, $S_{q,c}$), respectively, the number of qubits required to encode the combined system is $n = (2 n_\mathrm{c} + n_\mathrm{q})\log{N^\mathrm{semi}_\mathrm{grid}}$, where the factor of $2$ multiplying the classical particle number, $n_\mathrm{c}$, comes from the doubling of the Hilbert space of KvN systems. For this type of system, $N^\mathrm{semi}_\mathrm{grid} = S_q/S_c \;N^\mathrm{quant}_\mathrm{grid}$ where $N^\mathrm{quant}_\mathrm{grid}$ represents the number of grid points that would be necessary if no semiclassical approximation were made. Thus, for instance, the resource savings in terms of the ratio, $R$, of qubits used in a quantum-semiclassical simulation versus a fully quantum simulation for the encoding is 
\begin{equation}
  R = \frac{3}{2} \log_{N^{\mathrm{quant}}_{\mathrm{grid}}}\left(\frac{S_q}{S_c}N^\mathrm{quant}_\mathrm{grid} \right) \; ,
\end{equation}
assuming equal numbers of particles, $n_c = n_q$. Here, for instance for a mixed simulation of a proton-electron system, where $S_q/S_c \sim 10^{-3}$, $R \approx 0.6$, for a system of size $N^{\mathrm{quant}}_{\mathrm{grid}} = 10^{6}$. Hence, for this example, the advantage of using an MQS simulation, as opposed to a fully quantum simulation, is nominal since it takes just a few more qubits to resolve the full grid.

In the second case, consider a quantum simulation of two fields \cite{jordan2012quantum} (or, in general, a second-quantized system), where each individual grid point is represented by one quantum register per field (representing the occupation number of the quantum or classical field ($n_\mathrm{q,c \; field}$) at a given spatial location). Here, the resources for a fully quantum simulation would be
\begin{equation}
2\log(n_\mathrm{q\;field})\;N^\mathrm{quant}_\mathrm{grid} \; .
\end{equation}
And those for an MQS simulation would be
\begin{equation} 
 (\log(n_\mathrm{q\;field}) + 2\log(n_\mathrm{c\;field}))\;\frac{S_q}{S_c} N^\mathrm{quant}_\mathrm{grid} \; .
\end{equation}
From these expressions, we find that a quantum-semiclassical simulation will use
\begin{equation}
  R = 
\left(\frac{3}{2}\right)\;\frac{S_q}{S_c} 
\end{equation}
fewer qubits for field simulations, where we have set $n_\mathrm{q\;field} = n_\mathrm{c\;field}$.
In contradistinction with the situation for MQS particle simulations, MQS field simulations exhibit a reduction in qubit resources that is linear in the ratio $S_q/S_c$. 

One use-case where this is of particular value is semiclassical gravity. Semiclassical gravity (an MQS system) corresponds to summing all Feynman diagrams with no graviton loops (i.e. the semiclassical system), but an arbitrary number of matter loops (the quantum system) \cite{hu2008stochastic,anderson2003linear}.
For the example of MQS simulation of gravitational and scalar fields, we have the Einstein-Hilbert action
\begin{equation}
  S_\mathrm{EH} = -\frac{1}{16\pi G}\int d^4x \sqrt{-g}\;R
\end{equation}
and the action of a scalar field in a classical gravitational background
\begin{equation}
  S_\mathrm{scalar} = \frac{1}{2}\int d^4x \sqrt{-g}\;\left(g^{\mu\nu}\partial_\mu \phi \partial_\nu \phi + m^2 \phi^2 + \xi \phi^2 R\right)
\end{equation}
Here,
\begin{equation}
  G = 1/M_P = 1/1.22 \times10^{19} \mathrm{GeV}^{-1} \; ,
\end{equation}
with $M_P$ the Planck mass, giving $S_q/S_c \approx 4\times 10^{-18}$. 
It is clear that for this case, MQS simulations can give a significant advantage relative to quantum-quantum simulations. We note that for other MQS scenarios, such as Born-Oppenheimer, significant advantages can also exist \cite{brambilla2018born} with field-based MQS simulation.


\section{Applications}
\subsection{Theoretical Model}
\label{theoretical_model}
\begin{figure}[t!]
\centering 
\includegraphics[width=1\columnwidth]{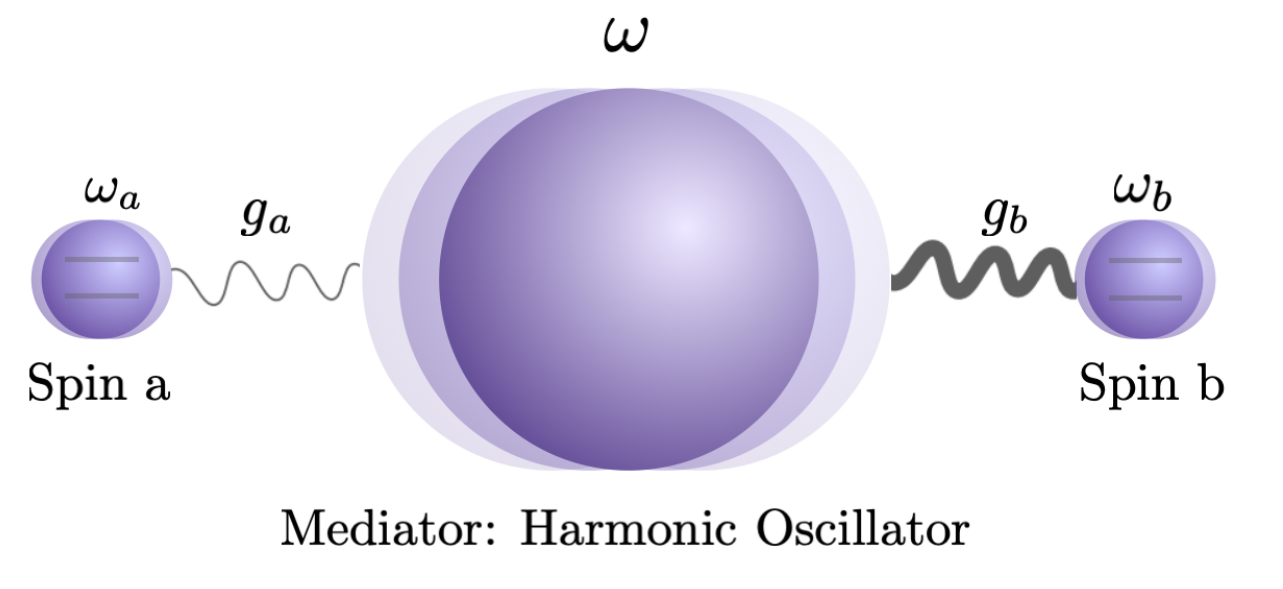}
\caption{Gravitational massive mediator between two quantum particles \cite{PhysRevLett.128.110401}. This model presents a system of two particles that interact through a massive mediator, which in the fully quantum case results in an  effective non-classical interaction between the two test-systems that grows with the mass of the mediator. On the other hand, when the mediator is a classical harmonic oscillator the effective non-classical interaction (measured by the logarithmic negativity) vanishes. See quantum simulation results in text.}
\label{fig:mental exp}
\end{figure}

\begin{figure*}
\centering 
\includegraphics[width=2.1\columnwidth]{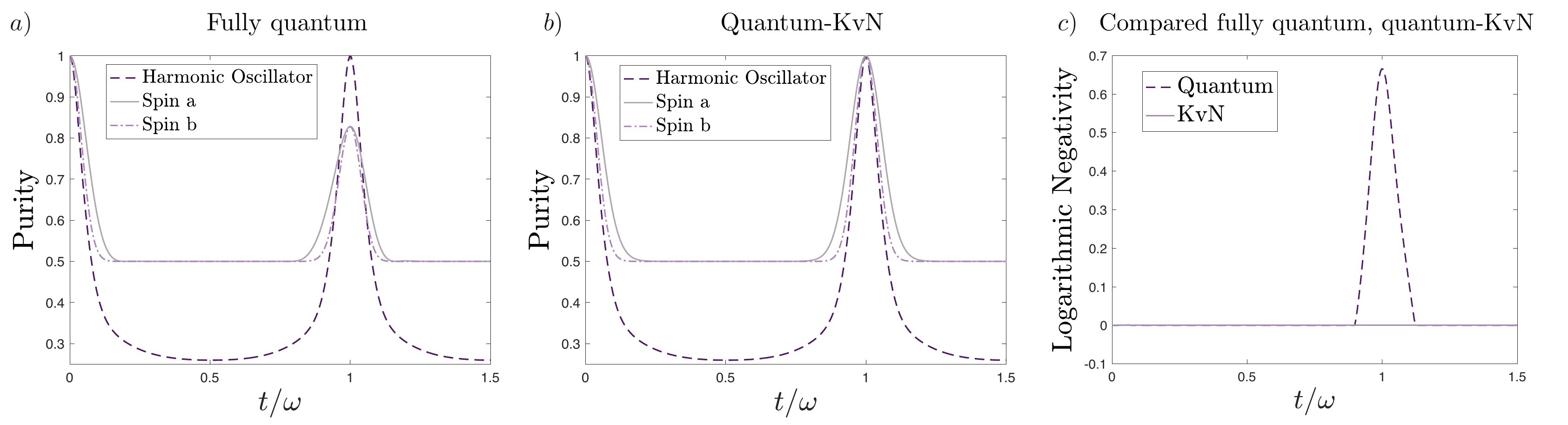}
\caption{Evolution of entanglement in quantum and MQS systems (Eqs. (\ref{eq:pedrnales}) and (\ref{eq:KvN_mediator}), resp.). We numerically simulated the dynamics of both quantum and classical oscillators with $n=7$ qubits for each coordinate, plus $1$ qubit for each spin subsystem, giving a total of $2 + 7 = 9$ qubits for the fully quantum case and $2 + 2\times 7 = 16$ qubits for the KvN model. a) The purity of spin and oscillator subsystems as a function of time for the fully quantum model (Eq.~(\ref{eq:pedrnales})). b) The purity of spin and oscillator subsystems as a function of time for the MQS model (Eq.~(\ref{eq:KvN_mediator})). c) The logarithmic negativity between the two spin subsystems in both quantum and MQS models. Note that in the MQS model, the interaction generates no distillable entanglement between the spin subsystems, whereas in the fully quantum model, the interaction generates significant entanglement.} 
\label{fig:num results}
\end{figure*}

Below, we apply our quantum algorithm to study a model from both fully quantum and MQS perspectives. We choose a toy model of special relevance to the study of gravity, a system comprised of a massive harmonic oscillator mediating the interaction between two quantum particles~\cite{PhysRevLett.128.110401}. When the oscillator is quantum the dynamics results in an  effective non-classical interaction between the two test-systems (spins) that grows with the mass of the mediator and is independent of its initial state and, therefore, its temperature. Under the assumptions given in the original proposal \cite{PhysRevLett.128.110401}, the Hamiltonian of this model reads:
\begin{multline}
    \hat{H}=\hbar\omega_a \hat{\sigma}^z_a+\hbar\omega_b\hat{\sigma}^z_b  + \hbar \omega \hat{a}^\dagger\hat{a}  \\ + \hbar (g_a \hat{\sigma}^z_a +  g_b     \hat{\sigma}^z_b)(\hat{a}+ \hat{a}^\dagger)
    \label{eq:pedrnales}
\end{multline}
with $\omega_a$, $\omega_b$ the frequencies of the spin subspaces, $\omega$ the frequency of the quantum harmonic oscillator, and $g_a$, $g_b$ the couplings between each spin and the mediator. For our quantum simulation, we replace the ladder operators $a$, $\hat{a}^\dagger$ with their associated position and momentum operators, $\hat{q}$ and  $\hat{p}$, obtaining
\begin{multline}
    \hat{H}=\hbar\omega_a \hat{\sigma}^z_a+\hbar\omega_b\hat{\sigma}^z_b  + \frac{m \omega^2}{2}  \hat{q}^2+ \frac{1}{2m}  \hat{p}^2 \\+ g_a\sqrt{2m\hbar\omega} \hat{\sigma}^z_a \hat{q}  +g_b \sqrt{2m\hbar\omega}    \hat{\sigma}^z_b \hat{q}
\end{multline}
with $m$ the mass of the oscillator.

In order to simplify the simulations of this model, we \textit{nondimensionalize} the model with the change of variables,
\begin{equation}
     \hat{\Pi}=\hat{q} \frac{ \sqrt{ m \omega}}{\sqrt{\hbar}}  \ \ \ \ \ \hat{\xi}=\hat{p}\frac{1}{\sqrt{\hbar m \omega}}
\end{equation}
obtaining the Hamiltonian
\begin{multline}
    \hat{H}=\hbar\omega_a \hat{\sigma}^z_a+\hbar\omega_b \hat{\sigma}^z_b  + \hbar \frac{\omega}{2} \left(\hat{\Pi}^2  +\hat{\xi}^2\right)\\+  \hbar\sqrt{2} \left(g_a \hat{\sigma}^z_a +g_b \hat{\sigma}^z_b\right) \hat{\Pi }.
    \label{eq:fully quantum}
\end{multline}
We will refer to the dynamics of this system as the fully quantum system and its analytical solution is given in \cite{PhysRevLett.128.110401}.

In the semiclassical regime, the massive mediator is taken to be classical. In the KvN formalism, Eq.~(\ref{eq:KvNHamiltonian}), the Hamiltonian of the system reads

\begin{multline}
    \hat{H}=\hbar\omega_a \hat{\sigma}^z_a+\hbar\omega_b \hat{\sigma}^z_b  + \hbar \frac{\omega}{2} \left(\hat{\xi}\hat{\lambda}_{\Pi}  - \hat{\Pi} \hat{\lambda}_{\xi}\right)\\-  \hbar\sqrt{2} \left(g_a \hat{\sigma}^z_a +g_b \hat{\sigma}^z_b\right) \hat{\lambda}_{\xi} \; .
        \label{eq:KvN_mediator}
\end{multline}
This system is an MQS system. Note that this Hamiltonian is composed entirely of quantum operators, $\hat{\sigma}_a^z$, $\hat{\sigma}_b^z$, $\hat{\xi}$, $\hat{\Pi}$, $\hat{\lambda}_\Pi$, and $\hat{\lambda}_\xi$. The primary difference is that position, $\hat\xi$, and momentum, $\hat\Pi$, operators commute for the semiclassical subsystem. Hence, Cor.~\ref{cor:OptimalMQSSim}, holds, and the system may be implemented with exponentially fewer resources on a quantum computer than an equivalent algorithm implemented on a classical computer.

In order to implement our simulations, for both quantum, Eq.~(\ref{eq:fully quantum}), and classical, Eq.~(\ref{eq:KvN_mediator}), models, we chose the initial state of the mediator to be a coherent state (for KvN, the distribution in phase space matching the Wigner transform of a coherent state) $\ket{\alpha}$, whose wavefunction reads 
\begin{equation}
\small
\psi^{(\alpha)}(x,0)=C \exp \Bigg( -\frac{1}{2}\left(\hat{\Pi}-\Re[\alpha]\right)^2+i\Im[\alpha]\hat{\Pi} \Bigg)
\normalsize
\end{equation}
where $C$ is a normalization factor. Its phase space distribution, given by the Wigner function, results in
\begin{equation}
    W(\hat{\Pi},\hat{\xi})=\frac{2}{\pi}e^{-[(\hat{\Pi}-\Pi_0)^2+(\hat{\xi}-\xi_0)^2]} \; ,
\end{equation}
with $\Pi_0=\sqrt{2}\text{Re}(\alpha)$ and $\xi_0=\sqrt{2}\text{Im}(\alpha)$. For simplicity we chose $\alpha \in \mathbb{R}$
.
Note that the operators $\hat{\Pi}$ and $\hat{\xi}$ of these distributions are independent and therefore the distribution can be factored as a product of its marginal distributions, which can be encoded in disjoint Hilbert spaces. 


\subsection{Numerical Simulations}
We numerically simulated the dynamics of both quantum and classical oscillators with $n=7$ qubits for each coordinate, plus $1$ qubit for each spin subsystem, giving a total of $2 + 7 = 9$ qubits for the fully quantum case and $2 + 2\times 7 = 16$ qubits for the KvN model. We adopted a Trotter method, \cite{Trotter1959OnTP,suzuki1976generalized,sornborger1999higher}, where we leverage the quantum Fourier transform to construct the momentum operators. During the system's evolution we tracked two physical quantities, the purity of each subsystem, and the logarithmic negativity between the spins where the mediator was traced out. The physical values of the frequencies and coupling used in the simulation are depicted in Tab.~\ref{tb:values}.

Purity and logarithmic negativity are both used to understand correlations between quantum systems. Of the two, purity is easier to compute on a quantum computer with just two copies of the system. However, the logarithmic negativity is an entanglement monotone and serves as an upper bound on the distillable entanglement, and therefore of interest for our results here, in that it can increase under mixing, when classical information is lost. 

In the case of fully quantum systems, it is expected that entanglement between quantum spins, upon tracing over any of the subsystems, and the mediator will generate decoherence and hence an initially pure state will lose its purity. Less expected, but seen and understood in the literature \cite{Bondar_2019}, MQS systems can also exhibit a similar loss of purity due to quantum-classical correlations.

\begin{table}[t!]
\small
\centering
\begin{tabular}{|c||c|c|c|c|c|}
\hline
 Parameter & $\omega / (2\pi)$ & $g_a / (2\pi)$  & $g_b / (2\pi)$  & $\omega_a / (2\pi)$  & $\omega_b/ (2\pi)$    \\
\hline
\hline
Value  & 1.59 Hz   & 1.59 Hz & 2.23 Hz & 7.0 e-6 Hz  & 3.34e-4  Hz\\
\hline
\end{tabular}
\caption{Physical values of the frequencies and coupling used in the simulation of Eqs. (\ref{eq:pedrnales}) and (\ref{eq:KvN_mediator}) }
\label{tb:values}
\end{table}

We present the results from simulating Eqs. (\ref{eq:pedrnales}) and (\ref{eq:KvN_mediator}) in Fig. \ref{fig:num results}. As we can see, during the evolution of both models, purity (a,b) and logarithmic negativity (c) evolve.
In the fully quantum case, Fig.~\ref{fig:num results}~a), once the first oscillation is completed, the oscillator is decoupled from the spins, and the spins become entangled, as we can see in the logarithmic negativity of the fully quantum system, Fig.~\ref{fig:num results}~c). On the other hand, in the KvN model, Fig.~\ref{fig:num results}~b), the classical harmonic oscillator purity evolves, indicating quantum-classical correlations. However, as may be seen via the logarithmic negativity, Fig.~\ref{fig:num results}~c),
these correlations do not generate distillable entanglement between the spins as expected of a classical mediator.

In both quantum and MQS systems, purities and logarithmic negativities evolved differently. This indicates that quantum-quantum and quantum-classical backreactions can usefully be studied with MQS simulation algorithms on quantum hardware.

\normalsize


\section{Conclusions}

 This article presents a novel quantum algorithmic approach for simulating MQS systems by combining Koopman-von Neumann (KvN) and standard quantum simulation methods. By combining semiclassical and quantum simulation techniques, MQS methods take advantage of resource reductions made possible with quantum algorithms for semiclassical simulation, and additionally provide a superpolynomial speedup relative to classical simulation methods.
 
 The KvN approach is suitable for MQS simulations in the Schr\"odinger picture, benefiting from the uniform encoding of both KvN and quantum Hamiltonians. The key distinction between KvN and fully quantum Hamiltonian simulation lies in the system variables' commutativity in the KvN (semiclassical) part of the simulation.

We have demonstrated that MQS simulation algorithms based on the KvN formalism can significantly reduce resource requirements when compared to equivalent classical algorithms. Additionally, MQS algorithms can outperform their equivalent quantum-quantum algorithms in terms of resource usage as long as the ratio of actions, $S_q/S_c$, of quantum and semiclassical systems is large, and can give a particular advantage in the case of simulations of quantum and semiclassical fields.

To validate the effectiveness of our approach, we implemented a proof-of-principle gravitational interaction model involving two two-level quantum systems with interaction mediated by a harmonic oscillator. By comparing measures of entanglement dynamics between quantum and classical oscillators, this model provided valuable insights into the two systems' entanglement properties, confirming the viability and usefulness of our approach.

Our work highlights the potential of the mixed KvN/quantum simulation approach for MQS simulation and paves the way for further advances in MQS simulations, offering a promising direction for optimizing resource utilization in quantum computational studies.

Our proposal establishes a general framework for simulating MQS systems with a focus on achieving more efficient MQS simulations relative to fully quantum simulations and establishing differences in the resulting behavior of entanglement due to backreaction in these systems. 

Finally, we note that recent research studying Born-Oppenheimer systems \cite{simon2023improved} takes a complementary approach, showing that KvN methods with phase-kickback from an electronic simulation may be used for computing thermodynamical properties in the Born-Oppenheimer approximation. This work outlines a semiclassical KvN algorithm and detailed computational complexity analysis for the computation of thermodynamical quantities from micro-canonical and canonical ensembles. In this work, the authors read out the thermodynamic state of the KvN system with an efficient quantum algorithm providing a direct (efficient) estimation of the free energy. Of interest in this study, is that the semiclassical KvN algorithm makes use of a many-particle encoding. Given the results shown here, this could potentially be further improved by converting the algorithm to a coarse-grained, second-quantized formalism \cite{kassal2011simulating}.

\begin{acknowledgments}
We thank Julen Padernales for extensive help with the spin-mediator model that we used for proof-of-principle. We thank Prof. Iñigo Egusquiza for discussions regarding Koopman-Von Neumann theory, Rolando Somma for discussions regarding resource differences in quantum-quantum and quantum-semiclassical systems, Michael Ragone and Joe Gibbs for helpful discussions. ATS acknowledges initial support by the U.S. Department of Energy through the Los Alamos National Laboratory, under the ASC Beyond Moore’s Law project when this project was conceived. Los Alamos National Laboratory is operated by Triad National Security, LLC, for the National Nuclear Security Administration of U.S. Department of Energy (Contract No. 89233218CNA000001). During the project's research phase, ATS was subsequently supported by the U.S. Department of Energy (DOE), Office of Science, Office of High Energy Physics, QuantISED program. JGC acknowledges financial support from OpenSuperQ+100 (Grant No. 101113946) of the EU Flagship on Quantum Technologies, as well as from the EU FET-Open project EPIQUS (Grant No. 899368), also from Project Grant No. PID2021-125823NA-I00 595 funded by MCIN/AEI/10.13039/501100011033 and by “ERDF A way of making Europe” and “ERDF Invest in your Future,”  Basque Government through Grant No.  IT1470-22 and the IKUR Strategy under the collaboration agreement between Ikerbasque Foundation and BCAM on behalf of the Department of Education of the Basque Government, as well as from and UPV/EHU Ph.D. Grant No. PIF20/276. \end{acknowledgments}

\bibliography{semiclassical,quantum}

\clearpage 
\end{document}